\documentclass[a4paper,11pt]{article}
\usepackage{pos,wrapfig}

\title{The CMS inner Tracker Endcap PiXel upgrade}

\author*[a]{Sascha Liechti}
\author[]{ on behalf of the CMS tracker group}
\manuallySeparateAuthors{}
\affiliation[a]{Universitaet Zuerich, Physik-Institut\\
  Winterthurerstrasse 190, 8057 Zuerich\\
  Switzerland}

\emailAdd{sascha.liechti@cern.ch}

\abstract{The High Luminosity Large Hadron Collider (HL-LHC) at CERN is expected to collide protons at a center-of-mass energy of 14 TeV and to reach the unprecedented peak instantaneous luminosity of $7.5 \times 10^{34} \text{cm}^{-2} \text{s}^{-1}$ with an average number of pileup vertices of 200. This will allow the ATLAS and CMS experiments to collect a data sample with an integrated luminosity up to $4000\,\mathrm{fb}^{-1}$ during the project's lifetime. To adapt to these harsh conditions, the CMS detector will be substantially upgraded before starting the HL-LHC, a plan known as CMS Phase-2 upgrade. The entire CMS silicon pixel detector (IT) will be replaced and the new detector will feature increased radiation hardness, higher granularity and capability to handle higher data rate and longer trigger latency. The upgraded IT will be composed of a barrel part, TBPX, as well as small and large forward disks, TFPX and TEPX. The TEPX detector has four large disks on each end, extending the coverage up to $\lvert \eta \rvert \approx 4.0$. In TEPX, the modules are arranged in five concentric rings. In this contribution the new TEPX detector will be presented, with particular focus on the mechanics design  and thermal performance, using finite element methods. Effects of the material choice for the cooling pipes and disk support structures connecting the modules with the CO$_2$ coolant are presented as well.}

\FullConference{%
  41st International Conference on High Energy physics - ICHEP2022\\
  6-13 July, 2022\\
  Bologna, Italy
}

\begin{document}
\maketitle

\section{Introduction}

The planned upgrade of the LHC at CERN (HL-LHC), colliding protons at a center of mass energy of 14 TeV, is expected to achieve an unprecedented peak instantaneous luminosity of $7.5 \times 10^{34} \text{cm}^{-2}\,\mathrm{s}^{-1}$ with an average number of pileup vertices of 200. 

The ATLAS \cite{ATLAS:2008xda} and CMS \cite{CMS:2008xjf} experiments are estimated to collect a data sample with an integrated luminosity up to  $4000\,\mathrm{fb}^{-1}$ during the project's lifetime. To adapt to the extreme conditions foreseen at the HL-LHC, the CMS detector needs to be substantially upgraded. The upgrade is known as the CMS Phase-2 upgrade \cite{CMS:2017lum, Reimers:2021vfs, Takahashi:2022rki, DelBurgo:2019vvoi}. As part of this upgrade the entire CMS silicon pixel detector (IT) will be replaced with the new detector featuring higher radiation hardness and granularity as well as the capability to handle higher data rates and a longer trigger latency.

According to the time schedule, the prototyping phase of the IT ends by the end of 2022, with a final assembly completed by the year 2027.

\section{TEPX design}
\label{sec: TEPX design}

The inner tracker endcap pixel detector will consist of four TEPX structures per end, each containing two double dees, resulting in a total of 16 double dees in the whole detector (Fig.~\ref{fig:  InnerTrackerLayout}). 

Each dee is populated with 44 modules, arranged in five concentric rings  (Fig.~\ref{fig: ModuleLayout}).

\begin{wrapfigure}{r}{0.478\linewidth}
  \includegraphics[width=\linewidth]{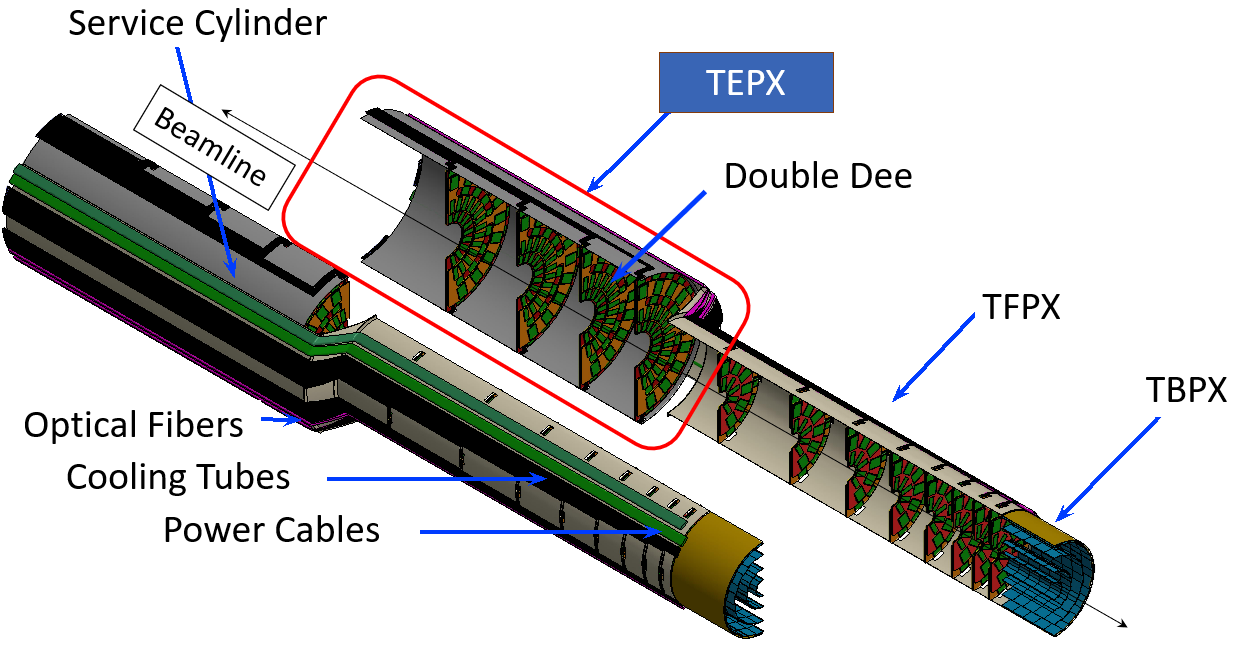}
	\caption{CMS IT layout of the Phase-2 upgrade.}
	\label{fig: InnerTrackerLayout}
\end{wrapfigure}

\subsection{Modules}

\begin{figure}[ht]
\begin{minipage}[b][][b]{0.55\linewidth}
\centering
         \includegraphics[width=0.89\linewidth]{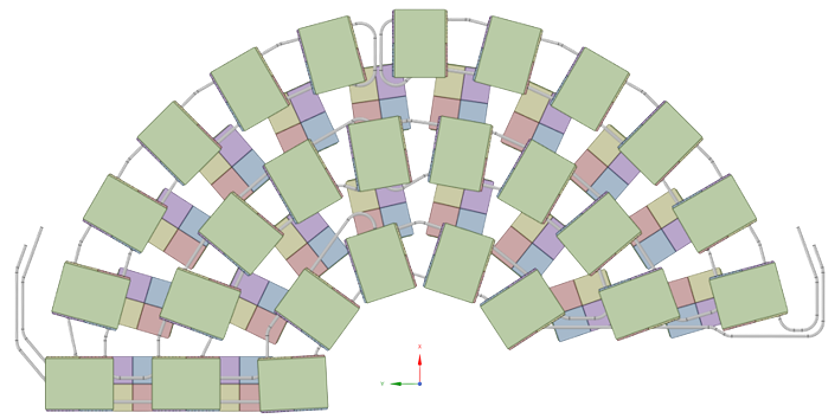}
         \caption{Placement of the modules on the dee with each subsequent ring being on the opposite side of the dee.}
         \label{fig: ModuleLayout}
\end{minipage}\hfill
\begin{minipage}[b][][b]{0.4\linewidth}
\centering
         \includegraphics[width=0.9\linewidth]{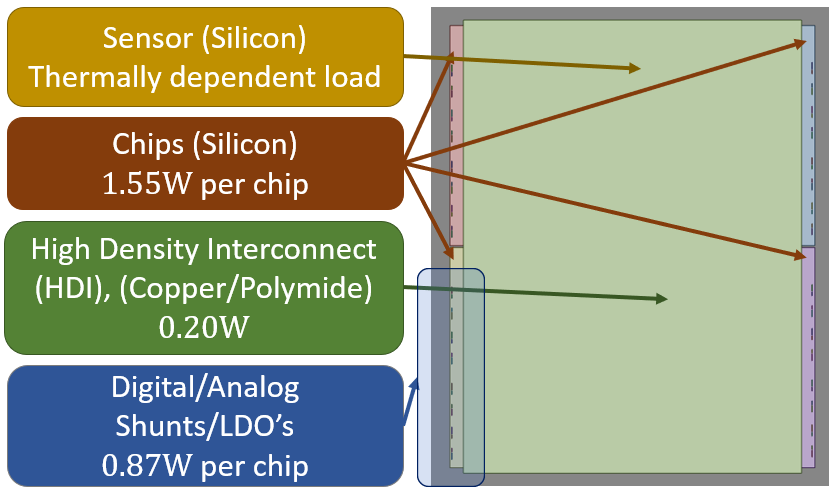}
         \caption{Design and thermal load and properties of a module.}
         \label{fig: ModelDesignAndThermalLoad}
\end{minipage}
\end{figure}

The module design, with its individual components and their thermal loads can be seen in Fig.~\ref{fig: ModelDesignAndThermalLoad}. 
A module consists of four readout chips, each equipped with four digital and four analog shunts and LDOs. The sensor is placed on top of the chips and is the only component with a temperature dependent thermal load. On top of the sensor is the high density interconnect, carrying the power and data from/to the modules. A module creates a thermal load of approximately 10W excluding the temperature dependent load of the sensor.

The thermal load of the sensor is exponentially dependent on their own temperature due to the leakage current in the sensor growing with temperature, in turn resulting in a higher power dissipation. The extreme case, where the increased thermal load cannot be completely absorbed by the cooling, results in the loss of thermal stability. We call this uncontrollable rise in temperature thermal runaway, which occurs once a critical temperature is reached.

The modules are cooled through cooling tubes within which a mixed phase CO$_2$ is circulated at $-35\,^\circ\mathrm{C}$. Using this CO$_2$ admixture allows for great cooling performance with little material needed, while simultaneously providing us with a small temperature gradient along the cooling pipe.

The readout electronics of the modules is located on the supply tubes, which provide mechanical support to the double dees. They are split into up to 12 distinctive electronics called portcards (Fig.~\ref{fig: TEPXSupplyTube}) per double dee. 

\subsection{Dee design}

\begin{figure}[ht]
\begin{minipage}[b][][b]{0.4\linewidth}
\centering
         \includegraphics[width=0.75\linewidth]{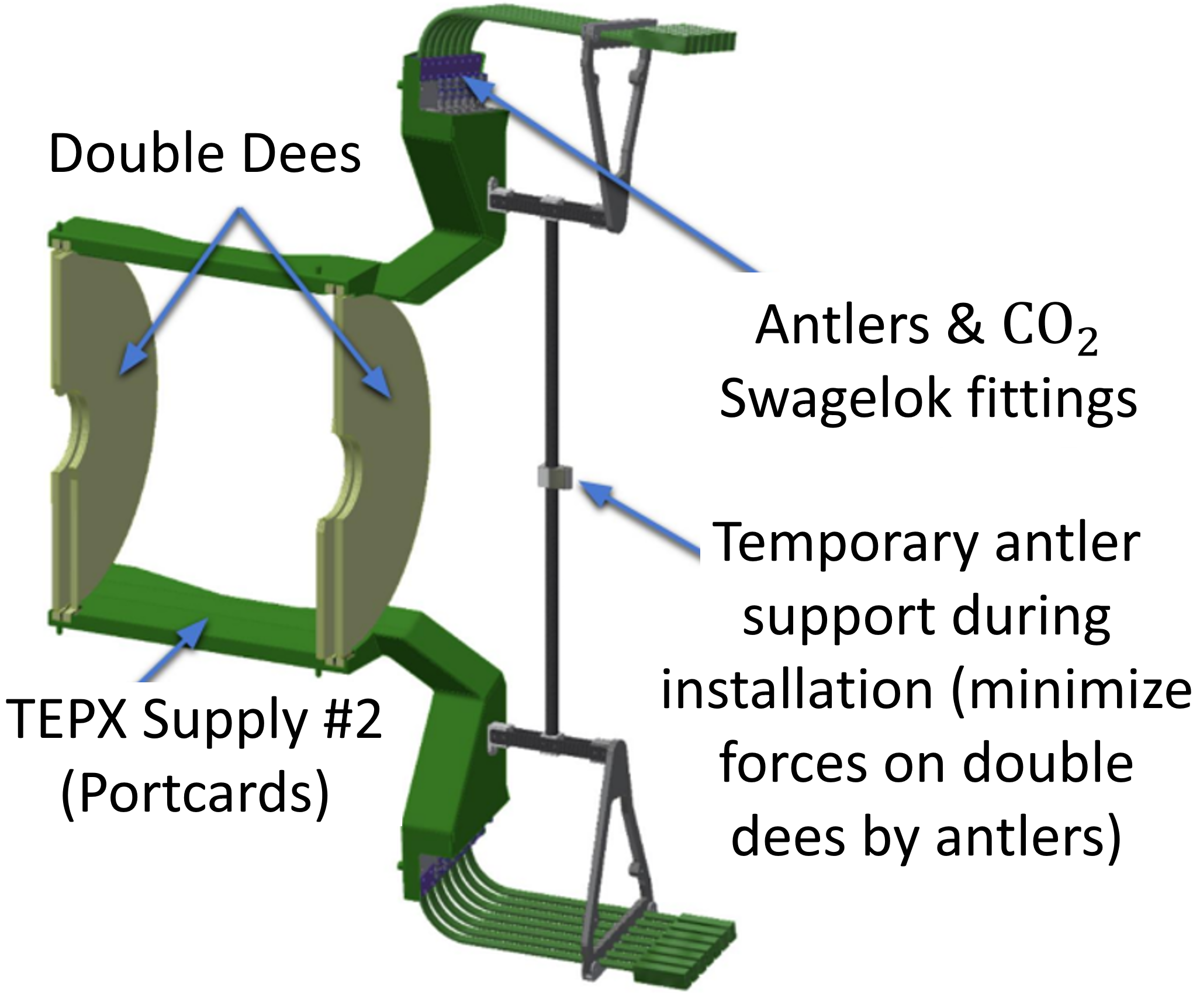}
         \caption{Layout of the TEPX supply tubes.}
         \label{fig: TEPXSupplyTube}
\end{minipage}\hfill
\begin{minipage}[b][][b]{0.58\linewidth}
\centering
         \includegraphics[width=0.9\linewidth]{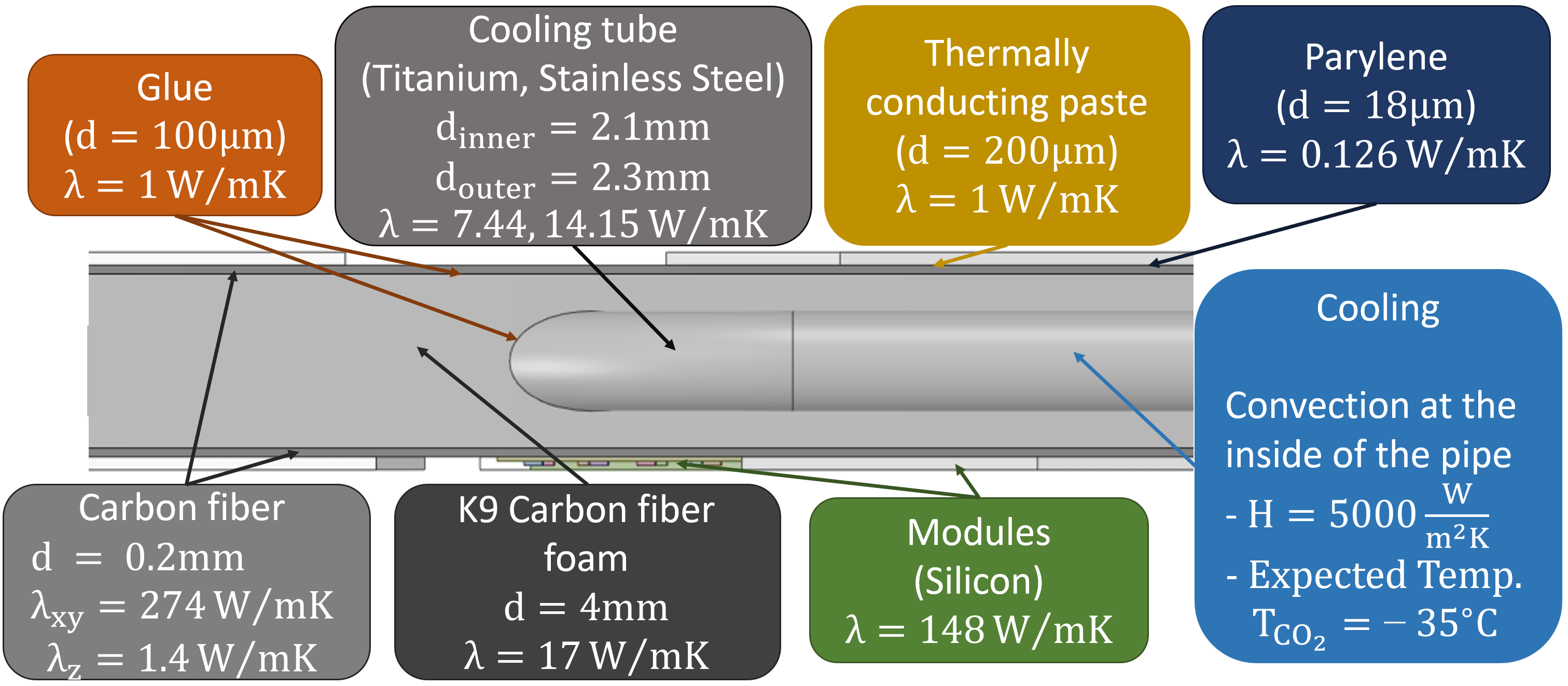}
         \caption{Design and thermal properties of a dee.}
		\label{fig: DeeDesignAndThermalLoad}
\end{minipage}    
\end{figure}

The whole dee is built with the goal of reducing weight and of increasing radiation length as much as possible while still providing adequate cooling. Looking at the transversal plane of the dee in Fig.~\ref{fig: DeeDesignAndThermalLoad}, we see the various of parts of the dee and their characteristics, like their thermal conductivity $\lambda$, their thickness d and the surface convection coefficient H between the cooling tubes and mixed phase CO$_2$. Thin cooling tubes run through the middle of the dee. These tubes, shown both in  Fig.~\ref{fig: ModuleLayout} and Fig.~\ref{fig: DeeDesignAndThermalLoad}, are made of stainless steel (SS). They are connected to highly thermally conductive carbon foam. Compressed carbon fiber plates (CKF) are placed on top of both sides of the carbon foam. They have a high thermal condictivity in the plane $\lambda_{xy}$, thus helping to spread the heat more evenly. The CFK and the carbon foam as well as the carbon foam and the cooling tubes are all connected via a highly thermally conductive glue. A thermal paste between the modules and the CFK ensures good thermal contact.  The other side of the modules is connected to a PCB (not shown in Fig.~\ref{fig: DeeDesignAndThermalLoad}), carrying power and data.

We plan to coat the modules with a thin parylene layer to insulate the read-out chips from the sensor as well as the chip backside from the CF of the dee. Also, we are exploring cooling tubes made out of titanium. Titanium provides approximately double the radiation length and half the density of SS, making it an appealing solution to the lightweight design. On the other side, Titanium has only half the thermal conductity of SS. This leaves us with 4 scenarios we want to analyze: For the first two scenarios we cover the modules in parylene and then vary the cooling tube material between SS and titanium. In the other two scenarios, we again vary the material of the cooling tubes between the two materials, however this time the modules are not coated in parylene.

\section{Results}

\begin{wrapfigure}{r}{0.44\linewidth}
	\centering
         \includegraphics[width=\linewidth]{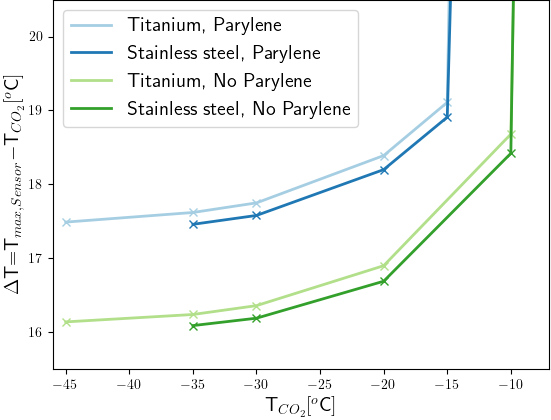}
         \caption{Sensor temperature as a function of the coolant temperature for all scenarios showing the thermal runaway behavior.}
         \label{fig: ThermalRunaway_K9-17}
\end{wrapfigure}

A finite element analysis using ANSYS \cite{ANSYS} was performed, exploring the four scenarios mentioned at the end of Sec.~\ref{sec: TEPX design},  simulating a fully populated dee, with 44 modules. The cooling was simulated under the assumption of no thermal gradient along the cooling pipe. We scanned the coolant temperature in steps of 5 $^\circ$C for all 4 scenarios, gradually increasing the coolant temperature until reaching the critical temperature.

Fig.~\ref{fig: ThermalRunaway_K9-17} shows that in all of the scenarios, we have a safe margin to thermal runaway at the expected operating temperature of the coolant of -35$^\circ$C. Adding a parylene coating to the modules results in an average +2$^\circ$C increase in temperature and insulates the read-out chip from the sensor and the chip backside from the CFK of the dee. Titanium tubes decrease the mass of the cooling tubes by half while maintaining the same thermal performance.

\end{document}